\documentclass[journal]{IEEEtran}
\usepackage{graphicx}
\usepackage{graphics} 
\usepackage{epsfig} 
\usepackage{float}
\usepackage{mathptmx} 
\usepackage{times} 
\usepackage{amsmath} 
\usepackage{amssymb}  
\usepackage[T1]{fontenc}
\usepackage{comment}
\usepackage{bm}
\usepackage{xcolor}
\usepackage[fitting]{tcolorbox}
\usepackage{cite}
\usepackage{hyperref}
\usepackage[utf8]{inputenc}
\usepackage[colorinlistoftodos]{todonotes}
\usepackage{multirow}

\begin{document}

\title{\LARGE \bf Enhancing Positronium Lifetime Imaging through Two-Component Reconstruction in Time-of-Flight Positron Emission Tomography
}
\author{Zhuo Chen, Chien-Min Kao, Hsin-Hsiung Huang, Lingling An 
\thanks{This work did not involve human subjects or animals in its research.}
\thanks{Z. Chen is with Department of Mathematics, University of Arizona, Tucson, AZ 85721 (e-mail: zchen1@math.arizona.edu).}
\thanks{C.-M. Kao is with Department of Radiology, University of Chicago, Chicago, IL 60637 (e-mail: ckao95@uchicago.edu).}
\thanks{H.-H. Huang is with Department of Statistics and Data Science, University of Central Florida, Orlando, FL 32816 (email: hsin.huang@ucf.edu)}
\thanks{L. An is with Department of Biosystems Engineering, University of Arizona, Tucson, AZ 85721 (email: anling@arizona.edu).}
\thanks{H.-H. Huang and L. An are co-corresponding authors of this work.}
}


\maketitle

\begin{abstract}
Positron Emission Tomography (PET) is a crucial tool in medical imaging, particularly for diagnosing diseases like cancer and Alzheimer's. The advent of Positronium Lifetime Imaging (PLI) has opened new avenues for assessing the tissue micro-environment, which is vital for early-stage disease detection. In this study, we introduce a two-component reconstruction model for PLI in Time-of-Flight (TOF) PET, incorporating both ortho-positronium and para-positronium decays. Our model enhances the accuracy of positronium imaging by providing a more detailed representation of the tissue environment. We conducted simulation studies to evaluate the performance of our model and compared it with existing single-component models. The results demonstrate the superiority of the two-component model in capturing the intricacies of the tissue micro-environment, thus paving the way for more precise and informative PET diagnostics.
\end{abstract}
\IEEEoverridecommandlockouts
\begin{keywords}
positronium lifetime imaging, image reconstruction, two-component positronium decay.
\end{keywords}
\IEEEpeerreviewmaketitle

\section{Introduction}
\begin{figure}[hb]
    \centering   \includegraphics[width=0.7\linewidth]{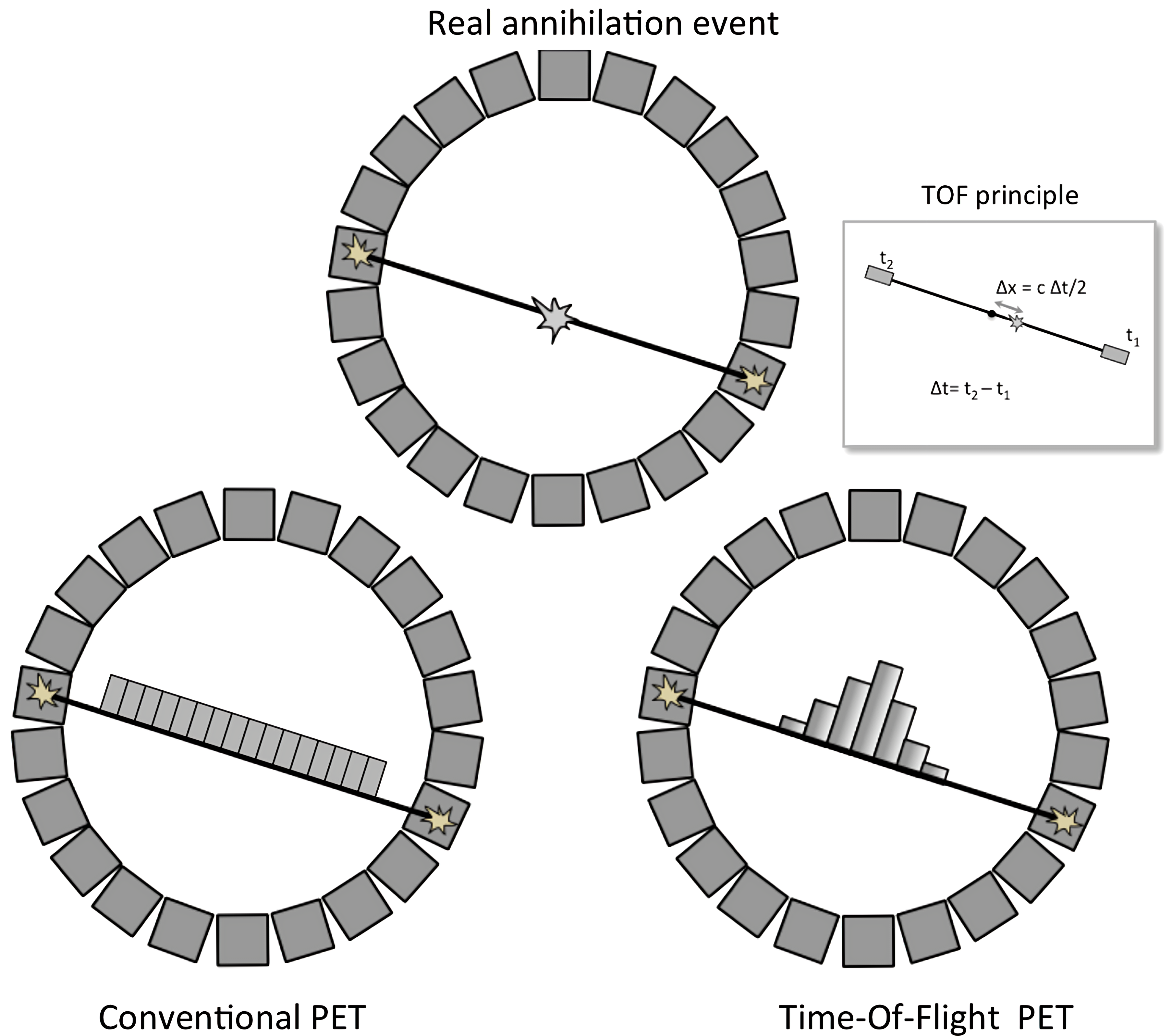}
    \caption{Illustration of a real positron annihilation event, conventional PET, and the time-of-flight PET \cite{recenttof_pet}.}
    \label{fig:pet_demo}
\end{figure}

Positron emission tomography (PET) is a medical imaging technique that has been extensively utilized in the diagnosis of diseases such as cancer and Alzheimer's disease \cite{pet_scan}. This imaging modality relies on the administration of molecules labeled with a positron emitting isotope that targets specific pathological signatures associated with these diseases. Upon a positron encounters an electron, a short-lived composition, i.e., positronium, is formed. As illustrated in Fig.~\ref{fig:pet_demo}, the positronium eventually gets annihilated, releasing two photons with an energy of 511 keV (these photons are frequently referred to as gamma rays) each in opposite directions \cite{pet_overview_shukla}. PET detectors capture these photons, and by analyzing their distribution, an image of the decay activity concentration in the body is created \cite{1990pet}. 

\begin{figure*}[t]
\centering
\includegraphics[width=.7\linewidth]{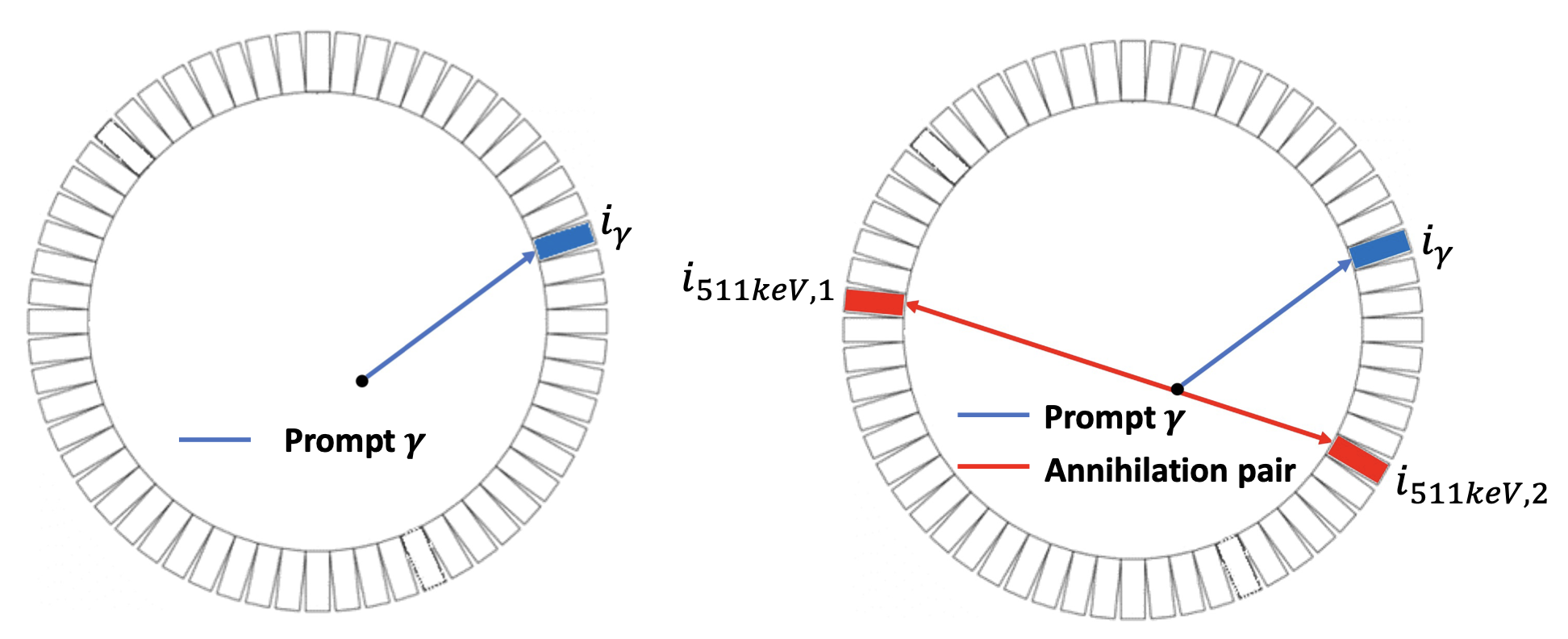}
\caption{A 2-D illustration of the emission and detection of prompt $\boldsymbol{\gamma}$ (in blue) and the annihilation pair (in red).}
\label{fig:pli_demo}
\end{figure*}
Conventional PET uses the detection time of the gamma rays only to identify the line along which the annihilation occurred (Fig.~\ref{fig:pet_demo}). It is unable, though, to determine which location along the line is the source of the two photons \cite{state_tof_pet}. Time-of-Flight (TOF) PET was introduced to  improves the temporal resolution of PET imaging \cite{recenttof_pet}, which uses the time-of-flight (the time it takes for the gamma rays to reach the detectors) difference of the gamma rays to better locate the annihilation position of the emitted positron (Fig.~\ref{fig:pet_demo}), allowing for more accurate localization of the annihilation event along the line of response (LOR, the line segment determined by the detector pair hit by the two gamma rays). While TOF PET has been successful in providing information about the decay activity concentration within the tissue, it has limited ability to directly assess the tissue micro-environment which carries important information such as tumor hypoxia \cite{moskalhypixia}.

Recently, positronium lifetime imaging (PLI) was invented \cite{proceeding2019,xx2,feasibility_2019} and the first positronium lifetime images were demonstrated ex-vivo using a novel multi-photon J-PET system \cite{x3_multiphoton}. Contrary to the traditional PET,
PLI uses an isotope such as Sc-44 that
emits a positron and a prompt gamma ray essentially at the same time\cite{matulewicz2021radioactive,choinski2021prospects}. Utilizing the emission time of the prompt gamma ray, PLI measures the lifespan of the positronium between the emission of the prompt gamma (in blue in Fig. \ref{fig:pli_demo}) and the annihilation of the positronium (in red in Fig. \ref{fig:pli_demo}) \cite{harpen,RevModPhys}. Positron annihilations can be decomposed into various components such as a long-lived ortho-positronium (o-Ps) and a short-lived parapositronium (p-Ps). o-Ps has a lifetime that is very sensitive to the size of the voids (free volume between the atoms) and the concentration in them of molecules such as oxygen molecules. As a result, it can provide information about the disease progression in an initial stage \cite{ops1,ops2}. It was shown in recent studies that the ortho-positronium lifetime in healthy adipose tissue differs from the ortho-positronium lifetime in cardiac myxoma tumors \cite{x4}. Knowing the ortho-positronium lifetime allows for a more complete understanding of physiological and biochemical interactions within the body.

Previous work in PLI reconstruction, including \cite{Qi2022, huang2023statistical, chen2023properties} focused on the lifetime of ortho-positronium, which has been found to be affected by the surrounding tissue environment, and only used o-Ps decays in simulated data. In this study, we delve into a two-components positronium decay model with the inclusion of a short-lived parapositronium and direct annihilation and adopted a two-component decay model, assuming known p-Ps rate-constant image, to reconstructed the rate-constant image of o-Ps decays.

\section{Methods}
By incorporating both long-lived ortho-positronium and short-lived parapositronium in the decay model, we assume that the lifetime of positronium follows a mixture of two exponentially modified Gaussian (EMG) distributions, one representing o-Ps decays and the other representing p-Ps decays, that is, the density of $\tau_k$ is expressed as
\begin{equation}
    w_1 \cdot \operatorname{EMG}\left(\tau_k ;  \lambda_{1,j}, \sigma^2\right)
+
w_2 \cdot \operatorname{EMG}\left(\tau_k ;  \lambda_{2,j}, \sigma^2\right)
\end{equation}
where $\operatorname{EMG}(\cdot)$ is the density of a EMG distribution and is given by
\begin{equation}
    \operatorname{EMG}\left(x ; \lambda_j, \sigma^2\right)=\frac{1}{2} \lambda_j e^{-\lambda_j\left(x-\frac{1}{2} \sigma^2 \lambda_j\right)}\left(1+\operatorname{erf}\left(\frac{x-\lambda_j \sigma^2}{\sqrt{2} \sigma}\right)\right) .
\end{equation}

The MLE of the o-Ps rate-constant image is given by
\begin{align*}
\footnotesize
& \arg \max _{\boldsymbol{\lambda_1}} \ell\left(\boldsymbol{\lambda}_{\mathbf{1}};  \boldsymbol{\lambda}_{\mathbf{2}}, \boldsymbol{w}_{\mathbf{1}} ,\hat{\boldsymbol{f}}, \mathcal{W}_{N_k}\right) =  
\arg \max _{\boldsymbol{\lambda_1}} \sum_{k=1}^{N_k} \log \\
&\left\{\sum_{j=1}^{N_j} H_{c_k, j} \hat{f}_j \left[w_1 \cdot \operatorname{EMG}\left(\tau_k ;  \lambda_{1,j}, \sigma^2\right)\right.\right.  \\
& +
\left.\left.w_2 \cdot \operatorname{EMG}\left(\tau_k ;  \lambda_{2,j}, \sigma^2\right)
\right]
\right\}
\end{align*}

Notations here follow \cite{huang2023statistical}, where
\begin{itemize}
    \item $\tau_k$: the lifetime of the $k$th positronium decay event between its emission and annihilation.
    \item $k = 1, \ldots, N_k$: the index of a positronium decay event.
    \item $N_k$: the total number of positronium decay events acquired.
    \item $\lambda_{1,j}$: the decay rate constant of o-Ps in the $j$th pixel. We refer to $\boldsymbol{\lambda_1}$ as the rate-constant image of o-Ps.
    \item $\lambda_{2,j}$: the decay rate constant of p-Ps in the $j$th pixel. We refer to $\boldsymbol{\lambda_2}$ as the rate-constant image of p-Ps.
    \item $j=1,\ldots,N_j$: the index of a pixel.
    \item $N_j$: the total number of images pixels.
    \item $\sigma$: variance of time measurements.  
    \item $f_j$: the number of decay events in the $j$th pixel. We refer to $\boldsymbol{f}$ as the activity image.
    \item $H_{c_k, j}$: an element of the system matrix $\boldsymbol{H}$ that is proportional to the probability that a positronium decay occurring inside image pixel $j$ would give rise to an event at line of response (LOR) $c_k$.
    \item $c_k$: The PET-TOF channel that detects the $k$th positronium decay event.
\end{itemize}

In addition, we introduced the weight image $\boldsymbol{w_1}$, which is the matrix representing the proportions of o-Ps among all decay events, and $\boldsymbol{w_2}$, the proportions of p-Ps among all decay events.

\begin{figure*}[t]
\centering
\includegraphics[width=.75\linewidth]{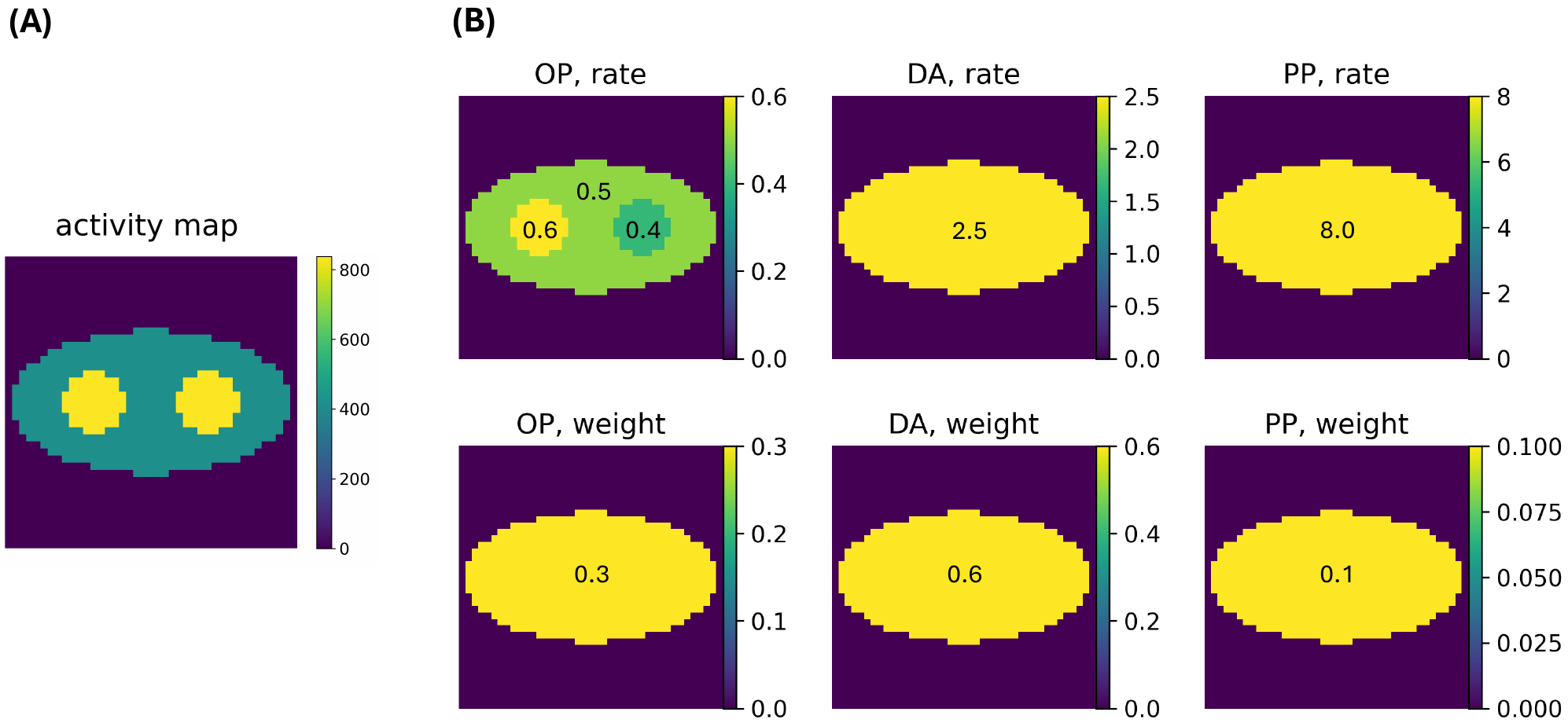}
\caption{(A) The ground truth of the activity map. (B) The ground truth of the rate-decay constant image and the weight image for o-Ps (OP), direct annihilation (DA), and p-Ps (PP).}
\label{fig:sim_setting}
\end{figure*}

PLI decay events were simulated using Monte Carlo methods described in \cite{huang2023statistical}. Three decay components, including o-Ps, p-Ps, direct annihilation (DA) were included when generating the decay events. Fig. \ref{fig:sim_setting} displays the ground truth of the activity image, and the rate-constant images and corresponding weight images for each component.

The MLE estimate of $\boldsymbol{f}$, denoted as $\hat{\boldsymbol{f}}$, is calculated using the maximum likelihood expectation maximization
(MLEM) algorithm \cite{ShopaDulski+2022+135+143}. The MLE estimate of $\boldsymbol{\lambda_1}$, denoted as $\hat{\boldsymbol{\lambda}}$ is calculated using the Limited-memory Broyden-Fletcher-Goldfarb-Shanno Bound (L-BFGS-B) method, due to its ability to deal with large numbers of variables and bound constraints. This optimization method is available from scipy.optimize in Python \cite{SciPy-NMeth2020}.

We implemented the proposed method using the combinations of the following simulation settings:

\begin{enumerate}
    \item[1)] $\boldsymbol{f}$ image used for $\boldsymbol{\lambda}$ reconstruction: 
    \begin{enumerate}
        \item [i)] true $\boldsymbol{f}$
        \item[ii)] estimated $\boldsymbol{f}$
    \end{enumerate}
    \item [2)] Timing uncertainty ($\sigma_t$)
    \begin{enumerate}
        \item [i)] 0.085 ns, the lowest timing uncertainty currently achieved
        \item [iii)] 0.242 ns, representing a 570 ps TOF PET system
    \end{enumerate}
    \item[3)] Event size, calculated as the product of the number of pixels in the ROI (627 in the phantom used) and a multiplier of choice which represents the number of PLI events per pixel.
    \begin{enumerate}
        \item [i)] $627\times100$
        \item [ii)] $627\times1000$
    \end{enumerate}
\end{enumerate}
For each simulation setting discussed below, 20 replications were conducted.

To evaluate the reconstruction performance, for each simulation, we calculated the normalized mean square error (NMSE) and the contrast recovery coefficient (CRC \cite{raczynski20203d}). In addition, we proposed a novel evaluation measure, the standardized absolute log ratio (SALR), which is defined as
\begin{equation}
    \text{SALR}_p=\frac{\left|\log \left(\frac{\hat{\boldsymbol{\lambda}}_{1, p}}{\hat{\boldsymbol{\lambda}}_b}\right)\right|}{\operatorname{SD}\left(\hat{\boldsymbol{\lambda}}_b\right) / \overline{\hat{\boldsymbol{\lambda}}_b}}, p=1,2
    \label{eqn:SALR}
\end{equation}
where $p=1,2$ represent the left and right circle in the phantom, and the subscript $b$ represent the background in the phantom. We evaluate SALR by taking the average of SALR$_1$ and SALR$_2$. Note that the denominator of Equation \ref{eqn:SALR} is the background variability of the reconstructed image as outlined in \cite{raczynski20203d}.

\begin{figure}[H]
\centering
\includegraphics[width=1\linewidth]{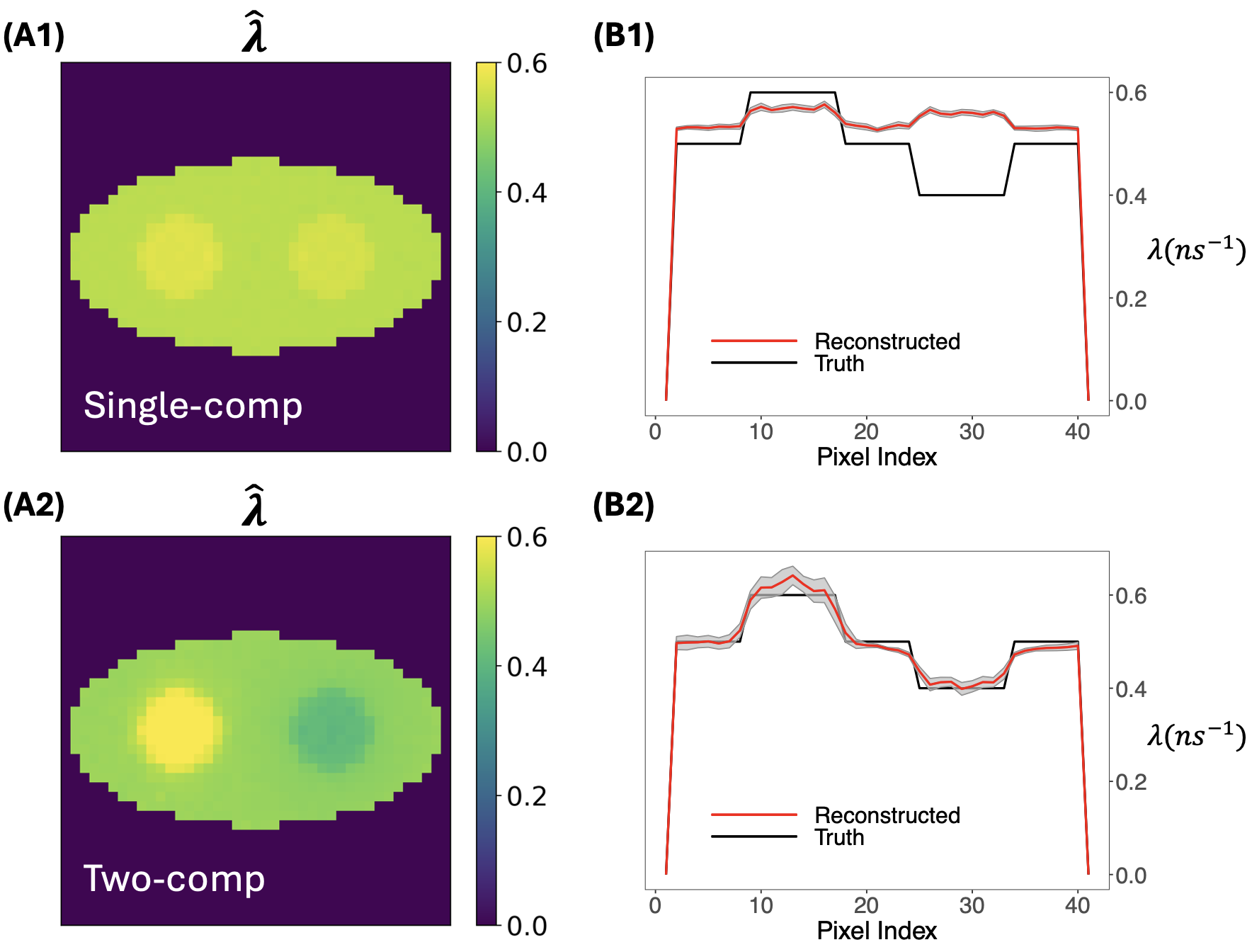}
\caption{(A1) The reconstructed rate-constant image using the single-component decay model. (B1) The horizontal profile across the center of the reconstructed image corresponding to the single-component decay model. (A2) The reconstructed rate-constant image using the two-component decay model.  (B2) The horizontal profile across the center of the reconstructed image corresponding to the two-component decay model.}
\label{fig:one_vs_two}
\end{figure}

\section{Results}

\subsection{Two-component decay model outperformed single-component decay model}
\begin{figure*}[t]
\centering
\includegraphics[width=0.7\linewidth]{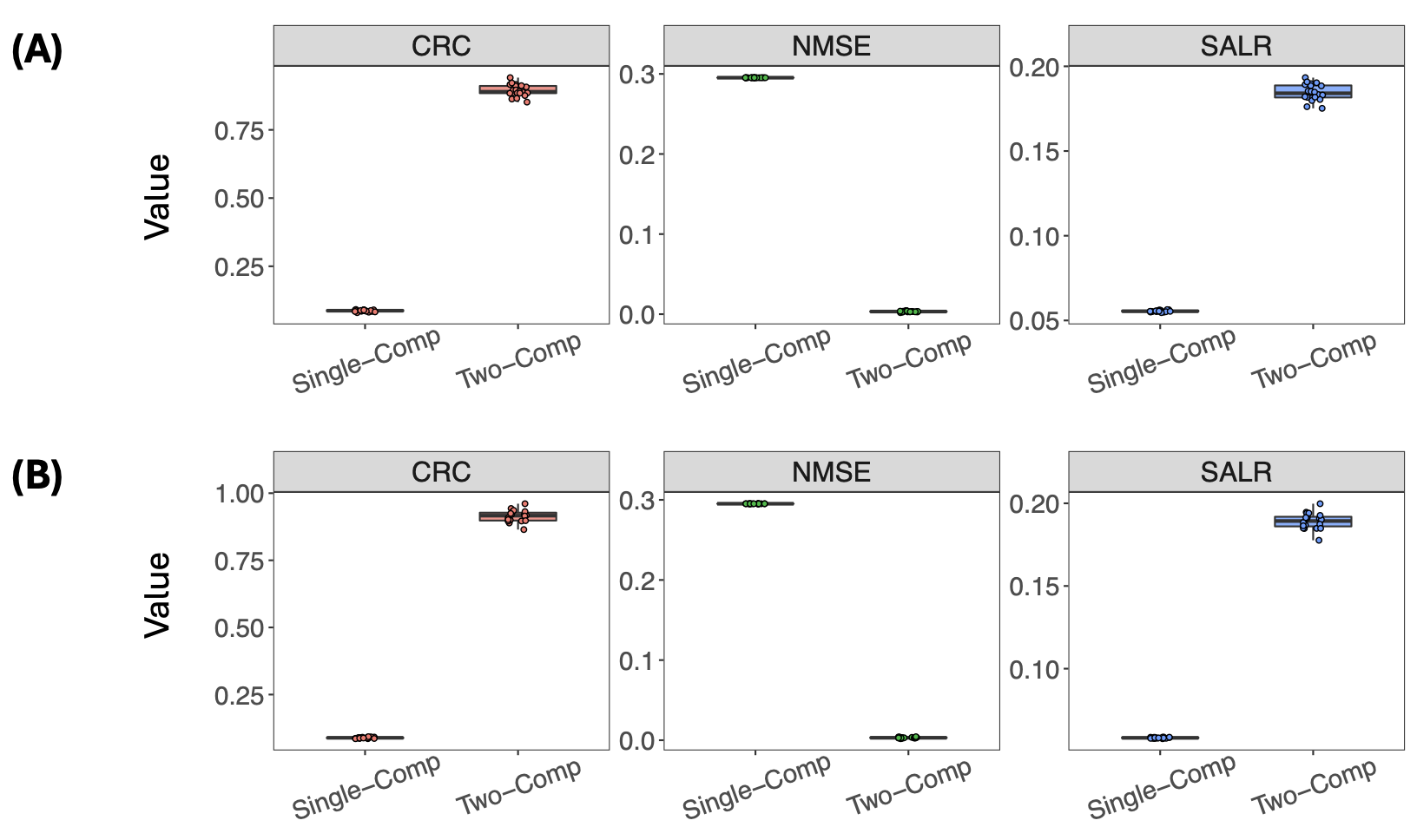}
\caption{(A) Box plots of CRC, NMSE, SALR, using estimated $\boldsymbol{f}$. (B) Box plots of CRC, NMSE, SALR, using true $\boldsymbol{f}$.}
\label{fig:one_vs_two_metrics}
\end{figure*} 

We first implemented the two-component model with the assumption that the weights and the rate-constant image of the short-lived decay component were known, for which we plugged in 2.5 ns$^{-1}$ as its decay rate-constant. Thus, the only unknown variable was the rate-constant image of the long-lived decay component, i.e., orth-positronium. Fig. ~\ref{fig:one_vs_two} compares the performance for the reconstruction of o-Ps rate-constant image using a two-component model vs. a single-component model. Fig.~\ref{fig:one_vs_two}(A1) and Fig.~\ref{fig:one_vs_two}(A2) are the averaged reconstructed image across the 20 replications. Fig.~\ref{fig:one_vs_two}(B1) and  Fig.~\ref{fig:one_vs_two}(B2) show the pixel values across the center row of the true and reconstructed image, referred to as the horizontal profile in this paper. The gray shaded region in the horizontal profiles represent the $\pm 1$ standard deviation range of the estimated image, calculated from the 20 replications. The reconstructed image of the two-component decay model closely aligned with the ground truth, while the single-component model failed to characterize the two circles in the phantom and had a nearly flat horizontal profile. The two-component decay model also shows a fairly low standard deviation, indicating high stability of the reconstruction results.

Fig.~\ref{fig:one_vs_two_metrics} compares the values of the three performance measures between the single-component and two-component decay model. Both the estimated and true $\boldsymbol{f}$ images were used to investigate the applicability of the proposed method in a realistic scenario where the truth $\boldsymbol{f}$ is not available. The two-component model demonstrated significantly superior performance in terms of all of the three measures. In addition, there was no significant difference in the performance measures between using the estimated $\boldsymbol{f}$ and true $\boldsymbol{f}$ for the two-component decay model, whereas the single-component model was more impacted by the use of the estimated $\boldsymbol{f}$.

\subsection{Two-component decay model is robust to unknown and constant weights}

Results shown above assumed that the weights of both o-Ps and p-Ps and the short-lived decay component are exactly know. We relaxed this assumption by introducing the unknown weight of o-Ps ($w_1$), and estimating the decay rate-constant image and the weight image of o-Ps simultaneously. Here, we assumed that the weight image of o-Ps is homogeneous, i.e., $\boldsymbol{w}_1 = [w_1]$ for all $j=1,\ldots, n_j$.

\begin{figure}[bh]
\centering
\includegraphics[width=1\linewidth]{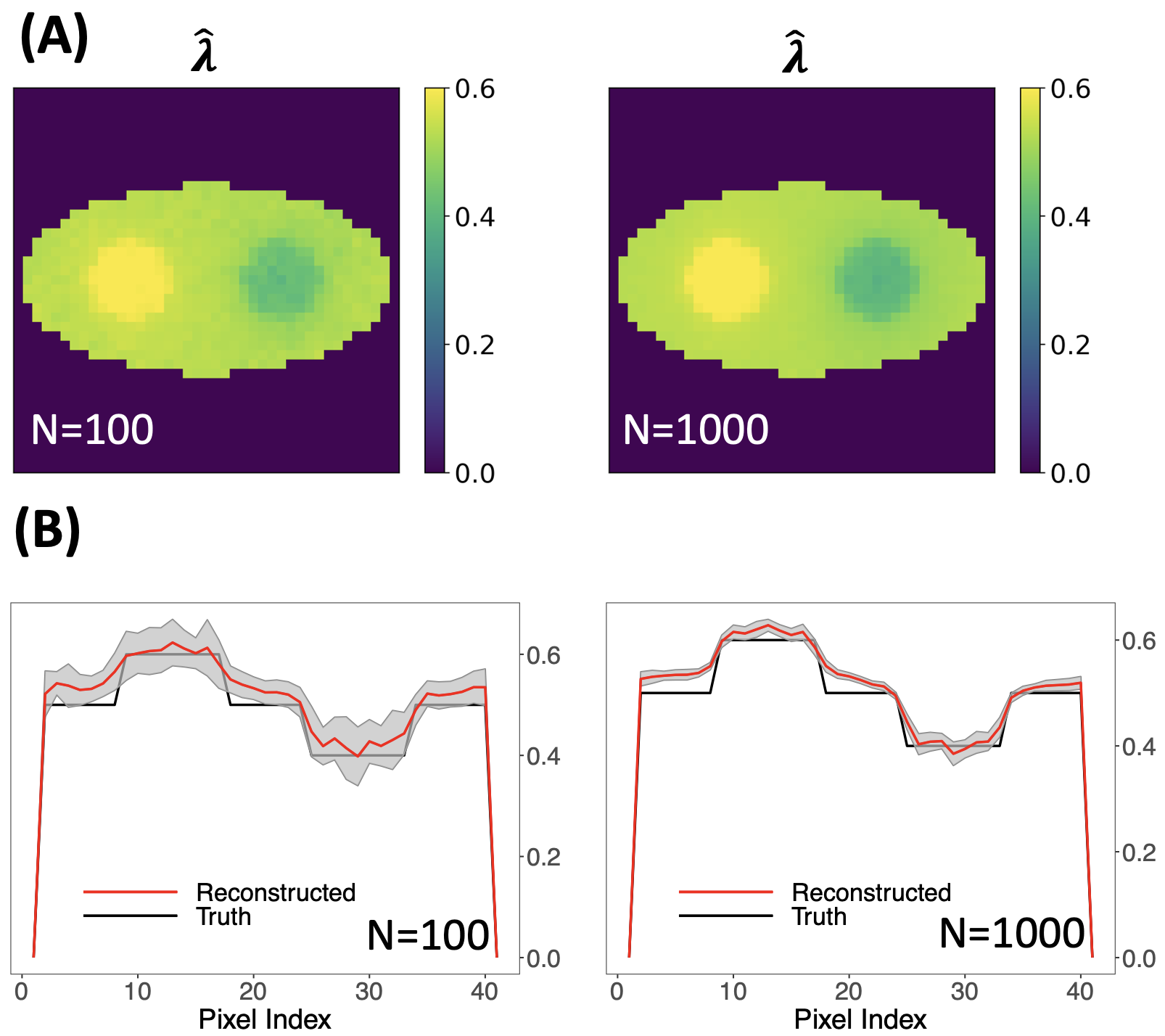}
\caption{(A) The reconstructed rate-constant images for o-Ps, using 100 and 1,000 events per pixel. (B) The horizontal profiles across the center of the reconstructed rate-constant image for o-Ps, using 100 and 1,000 events per pixel. }
\label{fig:events}
\end{figure}

Fig. \ref{fig:events}(A) shows the reconstructed rate-constant image of the two-component model, assuming that both $\boldsymbol{\lambda}_1$ and $\boldsymbol{w}$ are unknown. The number of decay events of both $627\cdot 100$ and $627\cdot 1000$ were used to test the performance of the method when only a low count of events are available. Fig. \ref{fig:events}(A) demonstrates that even when the number of decays is as low as 100 per pixel, the proposed method can still clearly capture the comparison between of the two circles vs. the background. 

Fig. \ref{fig:events}(B) exhibits the horizontal profile for the two event sizes used, together with the $\pm$1 SD range of the estimates. The horizontal profile for 100 pixels per pixel still aligned closely with the ground truth, with a higher variability in the reconstruction results compared to 1,000 events per pixel. These results demonstrated that the proposed method has great potential in generating highly quantitatively accurate rate-constant images for o-Ps with low-dose PLI imaging, with a slight compromise in the variability of the reconstruction.

\begin{figure}
\centering
\includegraphics[width=1\linewidth]{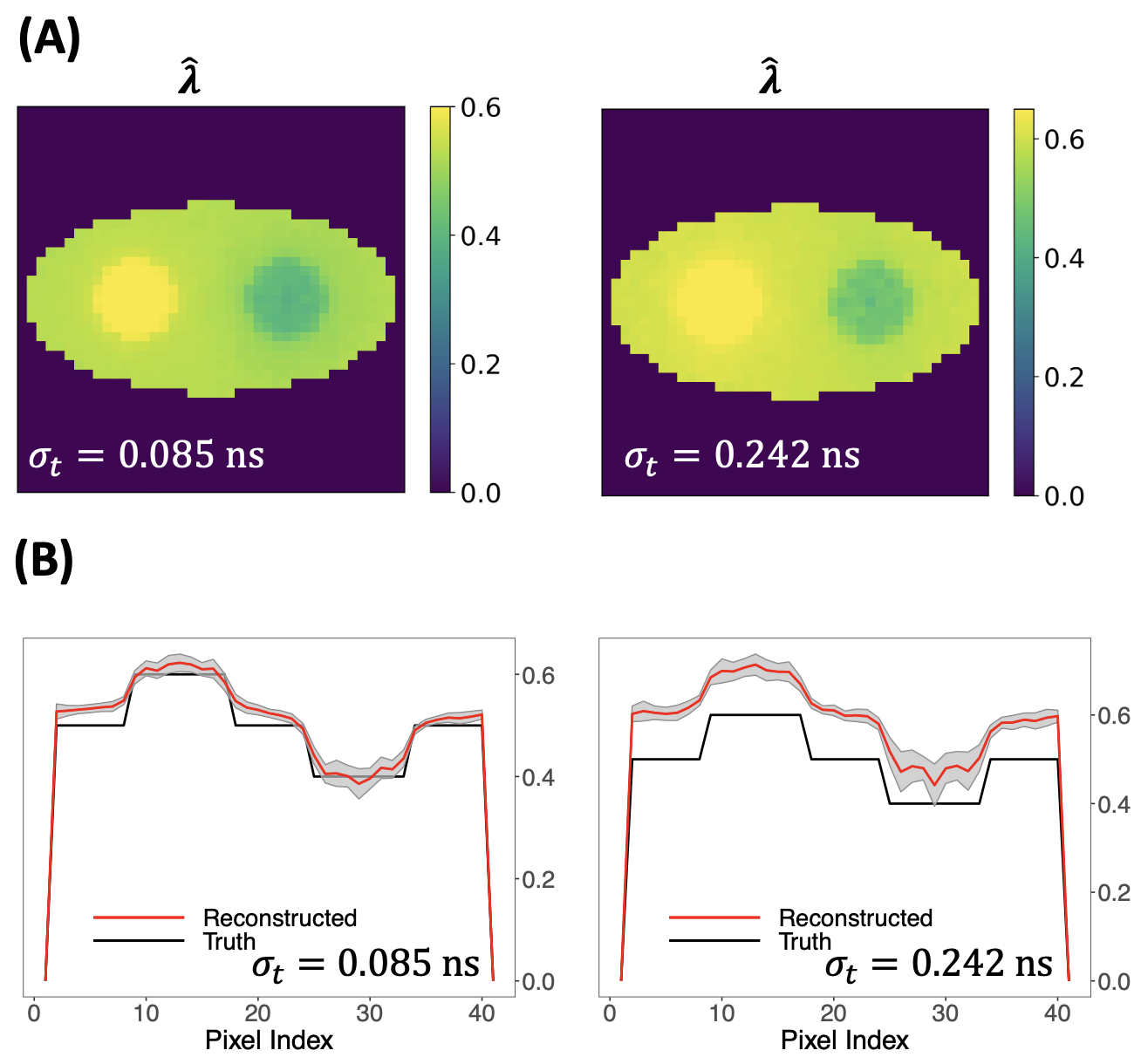}
\caption{(A) The reconstructed rate-constant images for o-Ps, using $\sigma_t=0.085$ ns and $\sigma_t=0.242$. (B) The horizontal profiles across the center of the reconstructed rate-constant image for o-Ps, using $\sigma_t=0.085$ ns and $\sigma_t=0.242$.}
\label{fig:sigma}
\end{figure}

In addition to event sizes, we assessed the proposed method's performance under different uncertainty in time measurements ($\sigma_t$). Two time resolution values, 0.085 ns and 0.242 ns were used. 0.085 ns is the lowest time uncertainty that can be achieved by the current TOF PET systems, while 0.242 ns is a typical time resolution, representing a 570 ps TOF PET system. Fig. \ref{fig:sigma} displays the reconstructed rate-constant images and horizontal profiles for the two $\sigma_t$ values. It shows that although $\sigma_t=0.242$ ns slightly overestimated the rate-constants compared to the truth, it still produced acceptable reconstructed image with fairly comparison of the two circle vs. the background, which demonstrates the robustness of our method against high time uncertainties of the TOF PET system. 

\subsection{Two-component decay model is robust to unknown and heterogeneous weights}

\begin{figure}
\centering
\includegraphics[width=1.0\linewidth]{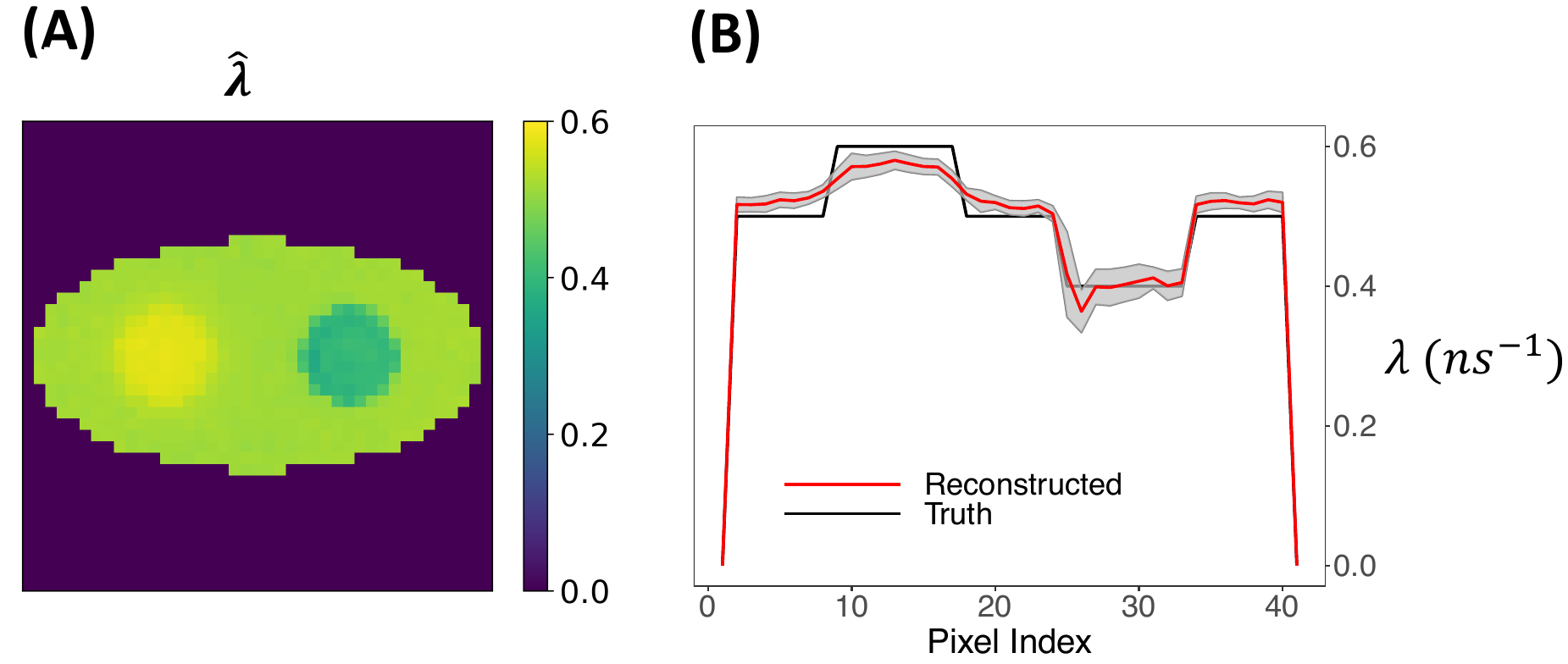}
\caption{(A) The reconstructed rate-constant image using a two-component model assuming heterogeneous weights. (B) The horizontal profile across the center of the rate-constant image using a two-component model assuming heterogeneous weights.}
\label{fig:hetero}
\end{figure}

Lastly, we further extended the proposed method to assuming that o-Ps could have various weights across its weight image, i.e., $\boldsymbol{w_1}=[w_j], j=1,\ldots,N_j$. This assumption introduced great complexity in parameter estimation due to the number of additional weights to be estimated. Thus, a higher event size of 2,000 pixels were used.

Fig. \ref{fig:hetero}(A) shows that the reconstructed rate-constant image maintained high contrast of the two circle vs. the background. In addition, the horizontal profile of the reconstructed image also closely match the ground truth, with a small $\pm$ 1 SD range for the 20 replications of the simulation.

\section{Conclusions}

 This study presents a significant advancement in positronium lifetime imaging through the development of a two-component reconstruction model in time-of-flight positron emission tomography. By incorporating both ortho-positronium and para-positronium decays, our model offers a more comprehensive understanding of the tissue micro-environment. The simulation studies conducted in this research clearly demonstrate the superior performance of the two-component model in capturing the intricacies of the tissue micro-environment compared to traditional single-component models. This highlights the potential of our model in enhancing the accuracy of PLI imaging, thereby contributing to more precise and informative medical imaging. The results of this study open up new possibilities for improving disease diagnosis and treatment planning, ultimately leading to better patient outcomes. Our work also lays the groundwork for further research in this area, which could lead to the development of even more advanced imaging techniques and technologies.

Future studies should explore the robustness and accuracy of the model under these conditions to ensure its applicability in low-dose imaging scenarios. Additionally, extending the model to include three decay components, which incorporates direct annihilation along with ortho-positronium and para-positronium decays, could provide a more comprehensive understanding of the tissue micro-environment. This extension would allow for a more detailed characterization of the decay processes, potentially leading to further improvements in the image quality of positronium lifetime imaging.

\bibliographystyle{IEEEtran}  
\bibliography{main}

\end{document}